\documentclass[preprint,showpacs,preprintnumbers,amsmath,amssymb,natbib]{revtex4}

\usepackage{graphicx}
\usepackage{epsfig}		
\usepackage{dcolumn}
\usepackage{bm}
\def\0{\mbox{\tiny $0$}}
\def\1{\mbox{\tiny $1$}}
\def\2{\mbox{\tiny $2$}}
\def\3{\mbox{\tiny $3$}}
\def\4{\mbox{\tiny $4$}}
\def\5{\mbox{\tiny $5$}}
\def\6{\mbox{\tiny $6$}}
\def\7{\mbox{\tiny $7$}}
\def\8{\mbox{\tiny $8$}}
\def\9{\mbox{\tiny $9$}}

\def\f14{\mbox{\tiny $\frac{1}{4}$}}

\def\L{\mbox{\tiny $L$}}

\def\R{\mbox{\tiny $R$}}

\def\F{\mbox{\tiny $F$}}

\def\D{\mbox{\tiny $D$}}

\def\sss{\mbox{\tiny $S$}}
\def\s{\mbox{\tiny $s$}}
\def\A{\mbox{\tiny $A$}}
\def\D{\mbox{\tiny $D$}}

\def\j{\mbox{\tiny $j$}}
\def\mi{\mbox{\tiny $-$}}

\def\pl{\mbox{\tiny $+$}}
\def\al{\mbox{\tiny $\alpha$}}

\def\bb#1{\mbox{\footnotesize $(#1)$}}

\begin{document}

\title{Coupling active and sterile neutrinos in the {\em cosmon} plus {\em seesaw} framework}

\author{A. E. Bernardini}
\affiliation{Departamento de F\'{\i}sica, Universidade Federal de S\~ao Carlos, PO Box 676, 13565-905, S\~ao Carlos, SP, Brasil}
\email{alexeb@ufscar.br, alexeb@ifi.unicamp.br}

\date{\today}

\begin{abstract}
The cosmological evolution of neutrino energy densities driven by cosmon-type field equations is introduced assuming that active and sterile neutrinos are intrinsically connected by cosmon fields through the {\em seesaw} mechanism.
Interpreting sterile neutrinos as dark matter adiabatically coupled with dark energy results in a natural decoupling of (active) mass varying neutrino (MaVaN) equations.
Identifying the dimensionless scale of the {\em seesaw} mechanism, $m/M$, with a power of the cosmological scale factor, $a$, allows for embedding the resulting masses into the generalized Chaplygin gas (GCG) scenario for the dark sector.
Without additional assumptions, our findings establish a precise connection among three distinct frameworks: the cosmon field dynamics for MaVaN's, the {\em seesaw} mechanism for dynamical mass generation and the GCG scenario.
Our results also corroborate with previous assertions that mass varying particles can be the right responsible for the stability issue and for the cosmic acceleration of the universe.
\end{abstract}

\date{\today}
\maketitle

The mass varying mechanism has often been suggested as a cosmological clock for the time evolution of the dynamical dark energy \cite{Ame07,Wet07}.
Theories that describe growing matter in such a context can possesses an adiabatic regime in which the quintessence field always sits at the minimum of its effective potential, which is set by the local mass varying dark matter density.
That is the prerogative of the so-called stationary condition of mass varying mechanisms \cite{Les06,Gu03,Far04}.
As a direct consequence of this exotic interaction, the relic particle mass is generated from the vacuum expectation value of the scalar field and becomes linked to its dynamics.
This coupled fluid is usually identified either as dark energy plus neutrinos, or dark energy plus dark matter \cite{Ber08A}.
In mass varying neutrino (MaVaN) scenarios, the time evolution of a cosmological scalar field can be interrupted by an increasing mass of the neutrinos.
Some approaches predict a transition from a cosmological scaling solution with dynamical dark energy at early times to a cosmological constant dominated universe at late times \cite{Wet07,Bro06A}.
In a particle physics realization of such a growing matter scenario, MaVaN's act as a trigger for the scalar field dynamics, giving an explanation to the ``why now problem'' \cite{Wet07,Pas09}.
The interface between MaVaN's and mass varying dark matter can also provide interesting relations among present values of neutrino masses, dark energy density, their corresponding equations of state and the cosmological stability conditions \cite{Ame07,Pas09}.

In this letter we provide some convincing constraints among MaVaN's, dynamical dark matter, and dark energy, through a single degree of freedom: a dynamical scalar field, $\phi\bb{a}$, or more explicitly, the cosmological scale factor, $a$.
Through a natural connection with the generalized Chaplygin gas (GCG) \cite{Kam02,Ber03}, we show that the intermediacy of the dark sector in
teractions done by $\phi\bb{a}$ can be exclusively relegated the scale factor, $a$, and correlated scales cosmologically driven by the GCG equation of state.
The first prerogative of our approach is that the cosmological evolution of active and sterile neutrino energy densities are driven by symmetrically coupled cosmon-type field equations \cite{Wet88}.
Active and sterile neutrino states are thus intrinsically connected by cosmon fields through the {\em seesaw} mechanism for mass generation \cite{seesaw}.
{\em Active} neutrinos presumably have tiny masses when compared to charged fermions, or to supermassive {\em sterile} neutrinos.
Such a smallness is understood in terms of the symmetry of the standard model (SM) of electroweak (EW) interactions amended by {\em seesaw} mechanisms.
As suggested by previous issues \cite{Ame07}, the {\em sterile} neutrino mass, $\mathcal{M}\equiv \mathcal{M}\bb{M\bb{\phi}}$, exhibits dynamical behaviour driven by the dark energy cosmon field, $\phi$.
Interpreting sterile neutrinos as the aforementioned dark matter adiabatically coupled with dark energy results
in active neutrinos with dynamical mass, i. e. MaVaN's.
They turn consistently into a perturbative effect driven by the abovementioned cosmon fields.
It results in a precise description of the dark sector energy components: dark energy, dark matter (as sterile neutrinos) and (active) neutrinos.
We consider that cosmon fields interconnect the dark sector and MaVaN components through the {\em seesaw} mechanism.

Dark matter is most often not considered in the formulation of the MaVaN models; however, the possibility of treating dark energy and dark matter in a unified scheme naturally offers this possibility.
Therefore, through the second prerogative of our approach, we assume that the mass varying behaviour, the cosmological dependence of the energy components, and the conditions for cosmological stability are all guaranteed by a quite interesting connection between the mechanism for mass generation and the GCG.
Its dependence with the scale factor indicates that the GCG can be interpreted as an entangled admixture of dark matter and dark energy.
We shall verify that identifying the dimensionless scale of the {\em seesaw} mechanism, $m/M$, with a power of the cosmological scale factor, $a$, allows for embedding the resulting dynamical masses into the GCG scenario for the dark sector.
Thinking on an effective model approach, it eliminates the intermediacy of dark sector interactions by a scalar field, in favor of simple scale dependencies.
Without additional constraints, our findings establish a precise connection among three distinct frameworks: the cosmon field dynamics, the {\em seesaw} mechanism for dynamical mass generation and the GCG equation of state.

The Lagrangian densities of active ($A$) and sterile ($S$) components in terms of matter fields, $\psi_{\A,\sss}$, can be written as
\begin{equation}
\mathcal{L}_{\A,\sss} = i\bar{\psi}_{\A,\sss}\gamma_{\mu}\partial^{\mu}\psi_{\A,\sss} + k_{\A,\sss}\bar{\psi}_{\A,\sss}\psi_{\A,\sss},
\label{aprl00A}
\end{equation}
where we have introduced two mass scales, $k_{\A} = \mu$ and $k_{\sss} = \mathcal{M}$, which respectively designate the tiny and large masses predicted by the {\em seesaw} mechanism,
\begin{equation}
-\mu = (M/2)[1 - \sqrt{1 + 4 (m/M)^{\2}}] ~~\mbox{and}~~\mathcal{M} = (M/2)[1 + \sqrt{1 + 4 (m/M)^{\2}}].
\label{aprl02}
\end{equation}
They correspond to the exact eigenvalues of the mass matrix $\left[\begin{array}{cc} 0 & m\\ m & M\end{array}\right]$ written in the orthogonal basis of chiral {\em left}- and {\em right}-handed neutrinos, $\nu_{\L,\R}$, related with the matter fields, $\psi_{\A,\sss}$, by
\begin{equation}
\psi_{\A} = \sqrt{s^{\2} + 1} (s \nu_{\L} - \nu_{\R})~~\mbox{and}~~
\psi_{\sss} = \sqrt{s^{\2} + 1} (\nu_{\L} + s \nu_{\R}),
\label{aprl00B}
\end{equation}
where we have introduced the the dimensionless scale $s = \sqrt{\mu/\mathcal{M}}$.
It results in $\mathcal{L}_{\A} + \mathcal{L}_{\sss} = \mathcal{L}_{\L} + \mathcal{L}_{\R} + \mathcal{L}_{\L\R}$, where $\mathcal{L}_{\L,\R}$ give the kinetic contribution and $\mathcal{L}_{\L\R}$ gives the mass mixing terms \cite{seesaw}.
The equivalence between the stress-energy tensor in chiral basis and matter field basis results in $\rho_{\A} + \rho_{\sss} = \rho_{\L} + \rho_{\R}$ and $p_{\A} + p_{\sss} = p_{\L} + p_{\R}$.
After introducing an auxiliary mass scale $\xi = \sqrt{\mu \mathcal{M}} \equiv m$, we provide the ingredients to define two energy scales, $\rho_{\1} = (\rho_{\sss} + \rho_{\A})/2$ and $\rho_{\2} = (\rho_{\sss} - \rho_{\A})/2$, which cosmologically evolve driven by symmetrically coupled {\em cosmon}-type equations.
In a FRW universe it corresponds to
\begin{equation}
\dot{\rho}_{\1,\2} + 3 H (\rho_{\1,\2} + p_{\1,\2}) - \dot{\phi}\frac{\mbox{d} \xi}{\mbox{d} \phi} \frac{\partial \rho_{\1,\2}}{\partial m} + \dot{\phi}\frac{\mbox{d} s}{\mbox{d} \phi} \frac{\partial \rho_{\2,\1}}{\partial s} = 0\\
\label{aprl02A}
\end{equation}
where $H = \dot{a}/{a}$ is the expansion rate of the universe, and the indices $1,\,2$ denote the symmetry under the interchange of $\rho_{\1}$ by $\rho_{\2}$.
The third term is the mass varying term and the last one is a mutual exchange term due to the coupling between states $1$ and $2$.
$\rho_{\A}$ and $\rho_{\sss}$ are evaluated like normal modes of $\rho_{\1}$ and $\rho_{\2}$ which are driven by symmetrically coupled cosmon-type equations.
$\rho_{\1}$ component can be identified in terms of matter fields while $\rho_{\2}$ is just an auxiliary energy scale that evolves with cosmon dynamics.

After combining Eqs.~(\ref{aprl02}), and observing that,
\begin{equation}
\rho_{\A}\bb{a, \xi, s} = \rho_{\A}\bb{a, \xi s}
\label{aprl023}
\end{equation}
and
\begin{equation}
\rho_{\sss}\bb{a, \xi, s} = \rho_{\sss}\bb{a, \xi/s}
\label{aprl023A}
\end{equation}
it is easy to identify the effective dynamical masses explicitly related to the cosmological evolution of active and sterile energy densities, $\rho_{\A}$ and $\rho_{\sss}$, by means of decoupled equations,
\begin{equation}
\dot{\rho_{\A}} + 3 H (\rho_{\A} + p_{\A}) - \dot{\phi}\frac{\mbox{d} \mu}{\mbox{d} \phi} \frac{\partial \rho_{\A}}{\partial \mu} = 0,
\label{aprl03A}
\end{equation}
and
\begin{equation}
\dot{\rho_{\sss}} + 3 H (\rho_{\sss} + p_{\sss}) - \dot{\phi}\frac{\mbox{d} \mathcal{M}}{\mbox{d} \phi} \frac{\partial \rho_{\sss}}{\partial \mathcal{M}} = 0.
\label{aprl03B}
\end{equation}
from where one notices that
\begin{equation}
\frac{\mbox{d} \ln{\xi}}{\mbox{d} \phi} =
(1/2)\left(\frac{\mbox{d} \ln{\mu}}{\mbox{d} \phi} + \frac{\mbox{d} \ln{\mathcal{M}}}{\mbox{d} \phi}\right)\nonumber\\
\end{equation}
and
\begin{equation}
\frac{\mbox{d} \ln{s}}{\mbox{d} \phi} =
(1/2)\left(\frac{\mbox{d} \ln{\mu}}{\mbox{d} \phi} - \frac{\mbox{d} \ln{\mathcal{M}}}{\mbox{d} \phi}\right)\nonumber\\
\end{equation}
The coupling between relic particles and the scalar field as described by Eqs.~(\ref{aprl03A})-(\ref{aprl03B}) are effective just for non-relativistic (NR) fluids.
Since the strength of the coupling is suppressed by the relativistic increase of pressure, as long as particles become ultra-relativistic (UR), with $T\bb{a} = T_{\0}/a \gg m\bb{\phi\bb{a}}$, the dark matter fluid, $\rho_{\sss}$, and the neutrino fluid, $\rho_{\A}$, decouple and evolve adiabatically \cite{Ame07}.
That is the case for ultra-relativistic (UR) regimes where $\frac{\partial\rho}{\partial m}\sim\frac{\partial\rho}{\partial s}\propto (\rho- 3 p)\approx 0$.
The mass varying mechanism expressed above translates the dependence of masses on the scalar field into a dynamical behaviour, i.e. $\mu\bb{a}$ and $\mathcal{M}\bb{a}$.

In some previous issues, we have already suggested that one could treat MaVaN's as a perturbative component derived from some previously unperturbed adiabatic solution $\rho_{\phi}$ \cite{Ber08A}.
The above results provide a precise explanation for this.
As one can observe, all the information from the dark sector (dark energy plus dark matter) acting on the (active) neutrino sector is carried out by the explicit dependence of $\mu\bb{\phi}$ on $\phi$.
From the cosmological point of view, it results in matter fields (active and sterile) that evolve separately.
It is no way the case for the coupled chiral eigenstates $\psi_{\L}$ and $\psi_{\R}$.
At primordial times, when $s^{\2} \sim 1$, such mass eigenstates should be indistinguishable, and the chiral eigenstates symmetrically defined.
At late times they turn into dark matter and (active) neutrinos, maintaining the correspondence among the flavour sectors.
To proceed, one can notice that the cosmon framework provides the right connection between dark matter (sterile neutrino) and dark energy through the well-known cosmon field equation,
\begin{equation}
\dot{\rho}_{\phi} + 3 H (\rho_{\phi} + p_{\phi}) + \dot{\phi}\frac{\mbox{d} \mathcal{M}}{\mbox{d} \phi} \frac{\partial \rho_{\sss}}{\partial \mathcal{M}} = 0.
\label{aprl03C}
\end{equation}
Eq.~(\ref{aprl03A}) for $i = A$ and Eq.~(\ref{aprl03C}) result in the adiabatic equation for the dark sector $(\rho_{\D\sss}, p_{\D\sss})$,
\begin{equation}
\dot{\rho}_{\D\sss} + 3 H (\rho_{\D\sss} + p_{\D\sss}) = 0,
\label{aprl04}
\end{equation}
with $\rho_{\D\sss} = \rho_{\phi} + \rho_{\sss} = \rho_{\phi} + \rho_{\1} + \rho_{\2}$ and $H^{\2} = \rho_{\D\sss}$ (with $H$ in units of $H_{\0}$ and $\rho_{\D\sss}$ in units of $\rho_{\mbox{\tiny Crit}} = 3 H^{\2}_{\0}/ 8 \pi G)$.
In spite of an intrinsic dependence on $\phi$, the equation of motion for the dark sector is not modified by $\rho_{\A}$.
The cosmological evolution of $\rho_{\A}$ can be perturbatively computed through Eq.~(\ref{aprl03A}).

The phenomenological consistency of this scenario can be quantified when one approximates
$\rho_{\A}$ and $\rho_{\sss}$ by the energy densities of a degenerate fermion gas (DFG) at different relativistic regimes,
\begin{eqnarray}
\rho_{\A}\bb{a} &=& (8 \pi^{\2})^{\mi\1}
\mu\bb{a}^{\4}\left[\eta\bb{a} (2 \eta\bb{a}^{\2} + 1)\sqrt{\eta\bb{a}^{\2} + 1} -
\mbox{arc}\sinh{(\eta\bb{a})}\right],\nonumber\\
\rho_{\sss}\bb{a} &=& (8 \pi^{\2})^{\mi\1}
\mu\bb{a}^{\4}\, s^{\mi\8}\left[\gamma s^{\2} \eta\bb{a} (2 \gamma^{\2} s^{\4} \eta\bb{a}^{\2} + 1)\sqrt{\gamma^{\2} s^{\4} \eta\bb{a}^{\2} + 1} -
\mbox{arc}\sinh{(\gamma s^{\2} \eta\bb{a})}\right],
\label{aprl04B}
\end{eqnarray}
where $\eta\bb{a} = q^{\F}_{\A}/(a \mu\bb{a})$,  $q^{\F}$ is the Fermi momentum, and the relation between the fluid thermodynamic regimes is parameterized by the coefficient $\gamma = q^{\F}_{\sss}/q^{\F}_{\A}$.
In particular, the DFG prescription is adequate for reproducing a consistent analytical transition between UR and NR regimes, and the inherent effects due to coupling dark matter ($\mathcal{M}\bb{\phi}$) and dark energy ($\phi$) can be verified through Eqs.~(\ref{aprl03A})-(\ref{aprl03C}).

\begin{figure}
\centerline{\psfig{file= 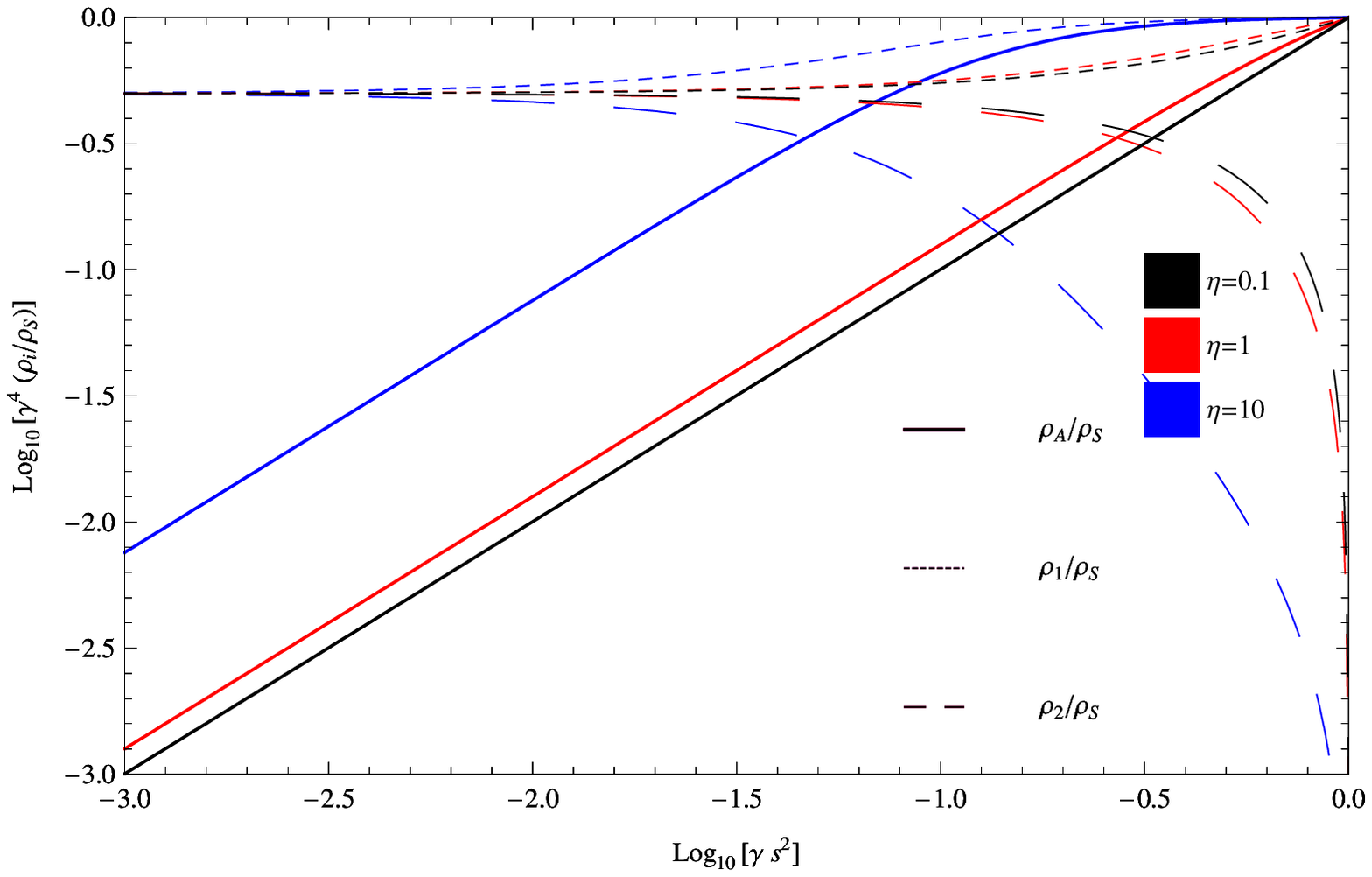,width=11 cm}}
\caption{\small
The exact correspondence between the energy density components and the ``modified'' {\em seesaw} scale parameter $\gamma s^{\2}$.
Sterile and active neutrinos are assumed to behave like a DFG.
For the NR limit, we have $\rho_{\A}/\rho_{\sss} \sim \gamma^{\mi\3} s^{\2}$.}
\label{PRL01}
\end{figure}

In the Fig.~\ref{PRL01} one observes the exact correspondence between the abovementioned energy density components and the ``modified'' scale parameter $\gamma s^{\2}$.
For the NR limit of a DFG, we have $\rho_{\A}/\rho_{\sss} \sim \gamma^{\mi\3} s^{\2}$.

The characteristic size of the active neutrino masses involves appropriate combinations of dimensionless (Yukawa) couplings $Y_{\j}$, i. e. $\mu_{\j} \sim m_{\j}^{\2}/\mathcal{M}$ with $m_{\j} \sim Y_{\j}\, v$ ($v\sim 2 \times 10^{\1\1}\,eV$), and the consistency with the observed oscillations requires for the mass of at least one neutrino $\mu_{\j}\gtrsim 0.05 \,eV$.
For $Y_{\j}$ of the order one this implies an upper bound for the large mass scale $\mathcal{M} \lesssim 10^{\2\3}\,eV$, which results in $s^{\2} \gtrsim 10^{\mi\2\4}$.
Once the present value of the rate $\rho_{\A}/\rho_{\sss}$ is phenomenologically assumed as $\mathcal{O}(10^{\mi\2})$, a DFG of (active) neutrinos at least approximately in the NR regime leads to $\gamma^{\mi\3}\sim 10^{\2\2}$, i. e. two completely different momentum scales for active and sterile neutrinos, $q_{\sss}/q_{\A} \sim 10^{\mi\7}$.
Sterile neutrinos behave like ultra cold dark matter.
The phenomenological value of $\rho_{\A}/\rho_{\sss}\sim\mathcal{O}(10^{\mi\2})$ is consistent with the $\Lambda$CDM cosmological model with (massive) cosmological neutrinos which transit from the UR to the NR regime in the CDM domination era \cite{Les06}.
The super massive sterile component is supposed to transit to the NR regime at energies much larger than $~1$ MeV, i. e.    earlier than the period where the relativistic particles: photons and electron/positrons were in equilibrium with neutrinos.
Thus, ultra cold dark matter could have become NR much earlier than neutrino decoupling from the SM cosmological plasma.
A basic understanding of the interaction rates of active neutrinos enables us to argue that active neutrinos were once kept in equilibrium with the rest of the cosmic plasma.
Ultra cold dark matter lost contact with the cosmic plasma slightly before reach the NR regime, so it did not inherit any remaining cosmic plasma associated energy.
The active neutrinos, then which did, are therefore hotter than the sterile ones.
They could have been "re-heated" by SM interactions which remain active after CDM decoupling as well as photons decoupled from neutrinos were re-heated due to electron-proton recombination.
It could result in the sterile/active momentum discrepancy given by $q_{\sss}/q_{\A} \sim 10^{\mi\7}$.
Once the partial derivatives of energy densities with respect to masses in Eqs.(\ref{aprl03A}-\ref{aprl03B}) can be written as
\begin{equation}
\frac{\partial \rho_{\A}}{\partial \mu} = \frac{ \rho_{\A} - 3  p_{\A}}{\mu},~~\mbox{and}~~
\frac{\partial \rho_{\sss}}{\partial \mathcal{M}} = \frac{ \rho_{\sss} - 3  p_{\sss}}{\mathcal{M}}
\label{aprl03ACC}
\end{equation}
they establish when the coupling to the scalar field is effective.
In particular, the condition given by Eq.~(\ref{aprl03ACC}) is guaranteed by the DFG approach.

In addition, if one assumes that dependence of the energy density on the scalar field could be parameterized by a $\lambda \phi^{\4}$ theory, or some type of quintessence potential the adiabatic evolution of the scalar field allows for obtaining the mass of the scalar field.
Such predictions for $m_{\phi}$ are consistent with the lower bound on the mass derived from assumptions that the scalar field has evolved adiabatically since nucleosynthesis.
The scalar field mass should then be greater than the Hubble constant at nucleosynthesis given by $\sim 10^{\mi\1\7}\, eV$.
However, the explanations for the adiabatic regime where the light scalar field have settled down to the minimum of its potential prior to nucleosynthesis are quite model dependent.
It requires a huge fine-tuning to generate the different scalar mass scales and it should be embedded in a more encompassing model, such as the Minimal Supersymmetric SM with the addition of one singlet chiral superfield \cite{Bal07}.
For a unified picture of dark energy and dark matter, the Higgs boson can be coupled to an additional singlet scalar field which we identify with a quintessence field \cite{Ber09}.
These predictions are consistent with the widely believed assertion that the SM of particle physics is actually only a low-energy approximation of a more fundamental underlying theory.
Such an interplay with cosmology represents an important guideline to obtain insights on the nature of a more fundamental theory.
It concerns, for instance, a natural connection with the cosmological scenario of the GCG.
The GCG model is characterized by an exotic equation of state \cite{Kam02,Ber03} given by
\begin{equation}
p = - A_{\s}  \rho_{\0} \left(\frac{\rho_{\0}}{\rho}\right)^{\al},
\label{aprl05}
\end{equation}
which can be obtained from a generalized Born-Infeld action \cite{Kam02}.
In any case, irrespective of its origin, several studies yield convincing evidence that the GCG scenario is a phenomenologically consistent approach to explain the accelerated expansion of the universe.
Notice that for $A_{\s} =0$, the GCG behaves always as matter whereas for $A_{\s} =1$, it behaves always as a cosmological constant.
This property makes the GCG model an interesting candidate for the unification of dark matter and dark energy, i. e. for the dark sector energy density, $\rho_{\D\sss}$, of our model.

Inserting the above equation of state into the unperturbed energy conservation Eq.~(\ref{aprl04}), one obtains through a straightforward integration \cite{Kam02},
\begin{equation}
\rho_{\D\sss} = \rho_{\0} \left[A_{\s} + \frac{(1-A_{\s})}{a^{\3(\1+\alpha)}}\right]^{\1/(\1 \pl \al)},
\label{aprl06}
\end{equation}
One of the most striking features of the above equations is that the GCG energy density interpolates between a dust dominated phase, $\rho_{\D\sss} \propto a^{-\3}$, in the past, and a de-Sitter phase, $\rho_{\D\sss} = -p_{\D\sss}$, at late times.
This evolution is controlled by the positive parameters, $\alpha$ and $A_{\s}$ with $0 < \alpha \leq 1$ .
Of course, $\alpha = 0$ corresponds to the $\Lambda$CDM model and we are assuming that the GCG model has an underlying real scalar field \cite{Kam02}.
All the phenomenological issues of the GCG have been recently addressed in Ref.\cite{Ber09}.

Assuming the canonical parametrization of $\rho_{\D\sss}$ and $p_{\D\sss}$ in terms of a scalar field $\varphi$,
\begin{equation}
\rho_{\D\sss} = \frac{1}{2}\dot{\varphi}^{\2} + V\bb{\varphi}, ~\mbox{and}~
p_{\D\sss}    = \frac{1}{2}\dot{\varphi}^{\2} - V\bb{\varphi},
\label{aprl07}
\end{equation}
and the Friedmann equation $H^{\2} = \rho_{\phi}$ in natural units with Planck mass $M_{Pl} = 1$ (i. e. with $H$ in units of $H_{\0}$ and $\rho_{\D\sss}$ in units of $\rho_{\mbox{\tiny Crit}} = 3 H^{\2}_{\0}/ 8 \pi G)$, one obtains the effective dependence of the scalar field $\varphi$ on $a$ given by
\begin{equation}
\varphi\bb{a} = - \frac{1}{2 \beta}
\ln{\left[\frac{\sqrt{1 + a^{\2\beta}A_{\s}/(1 - A_{\s})} - 1}{\sqrt{1 + a^{\2\beta}A_{\s}/(1 - A_{\s})} + 1}\right]}
\label{aprl08}
\end{equation}
where $\beta = 3(\alpha + 1)/2$.
From this point, the dynamics of the GCG scalar field is parameterized by Eq.~(\ref{aprl08}).
Thus we can consider the simplest version of the {\em seesaw} mechanism for which the Yukawa coupling between a light scalar field and a single neutrino flavour prescribes a linear dependence of the mass scale $m$ on $\varphi$, i. e. $m\bb{\varphi} \sim \varphi$.
By observing that the logarithm of the squared scale parameter $s^{\2}$ has an analytical structure similar to that of $\varphi\bb{a}$,
\begin{equation}
\ln{(s^{\2})} =\ln{(\mu/\mathcal{M})} = \ln{\left[\frac{\sqrt{1 + 4 (m/M)^{\2}} - 1}{\sqrt{1 + 4 (m/M)^{\2}} + 1}\right]}
\label{aprl09}
\end{equation}
we can rewrite the auxiliary scale $m/M$ in terms of the GCG parameters $A_{\s}$ and $\beta$, and of the scale factor $a$, as
\begin{equation}
\frac{m\bb{a}}{M\bb{a}} = \frac{m_{\0}}{M_{\0}}  a^{\beta} = \frac{1}{2} \sqrt{\frac{A_{\s}}{1-A_{\s}}} (\kappa \, a)^{\beta}
\label{aprl10}
\end{equation}
with the arbitrary constant $\kappa$ obviously introduced for phenomenological reasons ($\kappa$ disappears if one redefines $A_{\s}$ and $\rho_{\0}$ in Eq.~(\ref{aprl06})).
After simple mathematical manipulations, we obtain
\begin{equation}
\ln{(s^{\2})} =\ln{(\mu/\mathcal{M})} = -2\beta\varphi\bb{\kappa\,a},
\label{aprl09B}
\end{equation}
Adjusting the Yukawa coupling to the GCC scalar field parameters leads to $m\bb{\phi\bb{a}} = Y \phi\bb{a} \sim \varphi\bb{\kappa\,a}$.
Eqs.~(\ref{aprl09})-(\ref{aprl09B}) result in the following simplified forms for active and sterile neutrino masses,
\begin{equation}
\mu\bb{\varphi} = \varphi \exp{(-\beta \varphi)} ~~~~ \mbox{and} ~~~~ \mathcal{M}\bb{\varphi} = \varphi \exp{(+\beta \varphi)}.
\label{aprl11}
\end{equation}
where $\varphi\equiv\varphi\bb{\kappa\,a}$.
Since the scalar field constrained by the GCG dynamics decreases with $a$, for $\beta > 0$, active neutrino masses, $\mu$, increase with $a$, and sterile neutrino masses, $\mathcal{M}$, decreases with $a$.
Such a behaviour sets an obvious analytical dependence of the scale parameter, $s$, on $\varphi$ given by an exponential function: $s = \exp{(-\beta\varphi\bb{a})}$.
Such an exponential divergency between two mass scales is consistent with the naturalness of exponential potentials for several classes of quintessence models \cite{Wet07,Wet08}.
In addition, MaVaN's can be perturbatively implemented through Eq.~(\ref{aprl03A}).
As in some previous issues, growing neutrinos essentially stops the cosmological evolution of the scalar field and triggers an accelerated expansion of the universe.
It is evident that our findings are fairly general, as well as independent of the choice of the equation of state, and the form of $M\bb{\phi}$ and modifications on the equation of state for dark energy can lead to quite different scenarios.
For instance, it also admits static dark matter, which should result in neutrino masses given by $\mu\bb{a} \propto \exp{(-\beta \varphi\bb{\kappa\,a})} \equiv \exp{(-\beta \,Y\,\phi\bb{a})}$.
Turning back to the possibility of adiabatic instabilities, they were previously pointed out in cosmological scenarios in a context of MaVaN models for dark energy.
The essential ingredient of this class of models is that neutrino mass depends on the dynamical dark energy and grows in the course of the cosmological evolution.
The tiny neutrino mass and the recent accelerative era are twinned together through a scalar field coupling.
In the adiabatic regime, these models face catastrophic instabilities on small scales characterized by a negative squared speed of sound of the effective coupled fluid.
Such instabilities would give rise to exponential growth of small perturbations over the background uniform fluid \cite{Bro06A}.
The natural interpretation of this is that the Universe becomes inhomogeneous with neutrino overdensities subject to nonlinear fluctuations which eventually collapses into compact localized regions.

The mechanism that we have prescribed naturally leads to a fast increase of the neutrino masses, which results in a quite model dependent vanishing squared speed of sound at the present epoch.
The squared speed of sound is kept positive during the entire cosmological evolution
The GCG, as the cosmological background fluid, is consistent with all these results.
In particular, the dynamical mass is due to the time evolution of a cosmon field that coincides with the dynamics prescribed by a GCG universe.
These scenario is by no means the only possibility.
However, without any additional assumptions, our findings provide an attractive and convincing explanation for the why now problem through a confluence among three independent frameworks: the cosmon field dynamics, the {\em seesaw} mechanism for mass generation and the cosmological GCG scenario.
For active neutrinos, an increase of the mass $\mu$ by a factor $10^{\6}$ approximately corresponds to a decrease of the sterile neutrino mass $\mathcal{M}$ from the Planck scale to $10^{\1\3}\, GeV$.
It would largely be sufficient to load to dramatic consequences for cosmology, even in the present cosmological epoch.
As soon as neutrinos become non-relativistic, their coupling to the cosmon triggers an effective stop (or substantial slowing) of the evolution of the cosmon.
Naturally our results deserve a deeper analysis in which concerns the fine-tuning of the phenomenological masses involved.

The main features of our model can be summarized in the following terms.
i) The cosmological evolution of neutrino energy densities is driven by symmetrically coupled cosmon-type field equations that result in unmixed equations for active and sterile neutrino states intrinsically connected through the {\em seesaw} mechanism for mass generation.
ii) The possibility of treating dark energy and dark matter in the GCG unified scheme naturally couples dark matter with MaVaN's.
Interpreting sterile neutrinos as dark matter adiabatically coupled with dark energy provides the sufficient conditions to implement such unified picture in MaVaN formulation.
We have found the constraints imposed by the {\em seesaw} mechanism in order to establish the unique analytical connection to the canonical formulation of the GCG, in terms of a real scalar field.
iii) The dynamics of the mutual coupling among neutrinos, dark matter and dark energy are driven by only one degree of freedom: the scalar field, $\phi{a}$.
As is well known, the Higgs sector and the neutrino sector are possibly the only ones where one can couple a new standard model (SM) singlet without upsetting the known phenomenology.
Unfortunately, we know to much about the Higgs of the SM {\em seesaw} mechanism, i. e. the constraint on its interactions and properties.
It drops any pretension of classifying the generic scalar field of our model.
Since the GCG is formulated in terms of its equation of state that results in an explicit dependence on the scale factor, $a$, the introduction of the scales $s$ and $\xi$ allows for eliminating the intermediacy of the scalar field, in favor of assuming an explicit dependence on $a$, that is $\rho,\, p\bb{\phi\bb{a}}$ as $\rho,\, p\bb{a}$.
Due to the connection that we established between the GCG and the {\em seesaw} masses, our approach can also result in an effective model for MaVaN's coupling with the dark sector.
Eliminating the coupling action of the scalar field, in this case, the dynamics is driven by dimensionless scales $a$, $s\bb{a}$, and $\xi\bb{a}/M\bb{a}$.
To conclude, our unifying picture predicts the same features of standard quintessence models in mass varying matter scenarios which, however, falls through more ambitious unifying schemes. Our results reproduce the quintessence exponential laws besides corroborating with previous assertions that mass varying particles can be the right responsible for the stability issue and for the cosmic acceleration of the universe.
{\em Acknowledgments: This work was supported by the Brazilian Agencies FAPESP (grant 08/50671-0) and CNPq (grant 300627/2007-6).}

\pagebreak

\pagebreak
\newpage
\pagebreak
\begin{figure}
\centerline{\psfig{file= 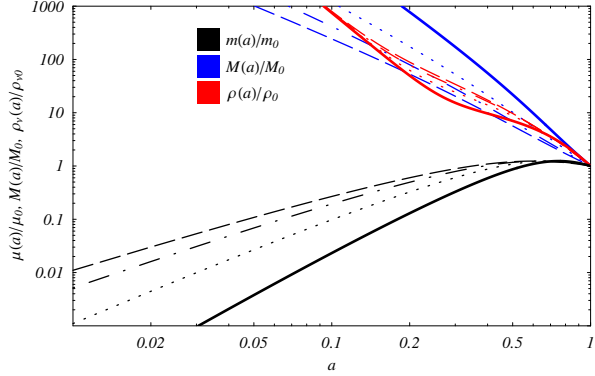,width=09 cm}}
\caption{\small Growing neutrino mass ($\mu\bb{a}/\mu_{\0}$), decreasing dark matter mass ($\mathcal{M}\bb{a}/\mathcal{M}_{\0}$) and neutrino energy density ($\rho_{\nu}\bb{a}/\rho_{\nu\0}$) in dependence on scale factor $a$.
They are respectively normalized by unitary quantities of $\mu_{\0}$, $\mathcal{M}_{\0}$ and $\rho_{\nu\0}$ at present ($a = 1$).
The results are obtained for different GCG scenario with $A_{\s}= 4/5$ and parameters $\alpha = 1$ (solid line), $1/2$ (dot line), $1/4$ (dash dot line), and $1/8$ (dash line).
}
\label{PRL02}
\end{figure}
\end{document}